\documentclass{aa} %
\usepackage{graphicx}
\usepackage{natbib}
\usepackage{txfonts}
\usepackage{epsfig}
\usepackage{longtable,lscape}

\def\harrow{\mathrel{\hbox{\rlap{\hbox{\raise4pt\hbox{${\rm +H}$}}}\hbox{$\longrightarrow$}}}}
\def\oarrow{\mathrel{\hbox{\rlap{\hbox{\raise4pt\hbox{${\rm +O}$}}}\hbox{$\longrightarrow$}}}}

\def\lesssim{\mathrel{\hbox{\rlap{\hbox{\lower4pt\hbox{$\sim$}}}\hbox{$<$}}}}
\def\gtsim{\mathrel{\hbox{\rlap{\hbox{\lower4pt\hbox{$\sim$}}}\hbox{$>$}}}}

\begin{document}
\title{The solar type protostar IRAS16293-2422: new constraints on the
  physical structure}

\author{Nicolas Crimier\inst{1} \and Cecilia Ceccarelli\inst{1} \and
  S\'ebastien Maret\inst{1} \and Sandrine Bottinelli\inst{2} \and
  Emmanuel Caux\inst{1} \and
  Claudine Kahane\inst{1} \and Dariusz C. Lis\inst{3} \and Johan
  Olofsson\inst{1} }

\offprints{N.~Crimier}

\institute{Laboratoire d'Astrophysique, Observatoire de Grenoble, 38041 Grenoble, France           
  \email{Nicolas.Crimier, Cecilia.Ceccarelli, Sebastien.Maret, Claudine.Kahane @obs.ujf-grenoble.fr}
  \and CESR-UPS, Centre National de la Recherche Scientifique, Toulouse, France \email{Sandrine.Bottinelli@cesr.fr}  
 \and California Institute of Technology, MC 301-17, Pasadena, CA 91125, USA \email{dcl@caltech.edu}}
%

%
%

 
  \abstract
  {The low mass protostar IRAS16293-2422 is a prototype Class 0
    source with respect to the studies of the chemical structure
    during the initial phases of life of Solar type stars.}
  {In order to derive an accurate chemical structure, 
    a precise determination of the source physical structure is
    required. The scope of the present work is the derivation of
    the structure of IRAS16293-2422.}
  {We have re-analyzed all available continuum data (single
    dish and interferometric, from millimeter to MIR) to derive accurate
    density and dust temperature profiles. Using ISO observations of
    water, we have also reconstructed the gas temperature profile.}
  {Our analysis shows that the envelope surrounding IRAS16293-2422 is
    well described by the Shu ``inside-out'' collapsing envelope model
    or a single power-law density profile with index equal to
    1.8. In contrast to some previous studies, our analysis does not
    show evidence of a large ($\geq 800$ AU in diameter) cavity.}
  {Although IRAS16293-2422 is a multiple system composed by two or
    three objects, our reconstruction will be useful to derive the
    chemical structure of the large cold envelope surrounding these
    objects and the warm component, treated here as a single source,
    from single-dish observations of molecular emission. 
}

   \titlerunning{The structure of the IRAS16293-2422 envelope}
   \maketitle
%


\section{Introduction}\label{sec:introduction}

Understanding how our Sun and Solar System formed is arguably one of
the major goals of modern astrophysics. Many different approaches
contribute to our understanding of the past history of the Solar
System. Analyzing the relics of the ancient eons, comets and
meteorites, is one. Studying present day objects similar to what the
Sun progenitor is another. Here we pursuit the latter approach and
analyze in detail the case of one of the best studied solar type
protostars, IRAS16293-2422 (hereinafter IRAS16293).  IRAS16293 is a
Class 0 protostar in the $\rho$ Ophiuchus complex at 120 pc from the
Sun \citep{Loi08} and has played the role of 
a prototypical solar-type protostar for astrochemical studies, just as
Orion~KL has done for high-mass protostars. This is beause of its
proximity and the resulting line strength of molecular emission
\citep[e.g.][to mention just
a few representative works from the previous
decades]{Wal86,Mun92,Bla94,Van95}. It is in this source that the
phenomenon of the ``super-deuteration''\footnote{The super-deuteration
  refers to the exceptionally high abundance ratio of D-bearing
  molecules with respect to their H-bearing isotopologues found in low
  mass protostars, with observed D-molecule/H-molecule ratios reaching
  the unity \citep[see e.g. the review in][]{Cec07}.}  has been first
discovered, with the detection of surprising abundant multiply
deuterated molecules: formaldehyde, hydrogen sulfide, methanol and
water \citep{Cec98,Vas03,Par03,But07}. It is also in this source that the
first hot corino \citep{Bot04,Cec07} has been discovered, with the detection of
several abundant complex organic molecules in the region where the
dust grain mantles sublimate \citep{Cec00a,Caz03,Bot04}.

Not surprisingly, therefore, IRAS16293 has been the target of several
studies to reconstruct its physical structure, namely the dust and gas
density and temperature profiles \citep{Cec00b,Sch02,Sch04,Jor05}, the
mandatory first step to correctly evaluate the abundance of molecular
species across the envelope. \citet{Cec00b} used water and oxygen
lines observations obtained with the Infrared Space Observatory (ISO)
to derive the gas and dust density and temperature
profile. Conversely, \citet{Sch02} used the dust continuum
observations to derive the structure of the envelope. Furthermore,
while Ceccarelli et al. assumed the semi-analytical solution by Shu \&
co-workers \citep{Shu77,Ada86} to fit the observational data,
\citet{Sch02} assumed a single power law for the density distribution,
and a posteriori verified that the Shu's solution also reproduced the
observational data. The two methods lead to similar general
conclusions: a) the envelope of IRAS16293 is centrally peaked and with
a density distribution in overall agreement with the inside-out
collapse picture \citep[][]{Shu77}; b) there is a region, about 300 AU in
diameter, where the dust mantles sublimate (giving rise to the
phenomenon of the hot corino, mentioned above). However, the two
different methods, not surprisingly, also lead to some notable
differences.  For example, the gas density differs by about a factor 3
in the region where the ice sublimation is predicted to occur, which
leads to differences in the derived abundances of several molecular
species.

Subsequent studies have built on the early ones to improve the
derivation of the physical structure of IRAS16293. First, the study by
\citet{Sch04}, based on interferometric OVRO observations,
concluded that the envelope has a large central cavity, about 800 AU
in diameter. Then, the work by \citet{Jor05}, using new
SPITZER data, concluded that the envelope has an even larger
central cavity, about 1200 AU in diameter. Such a large central cavity
has a great impact on the whole interpretation of the hot
corino of IRAS16293, as it predicts the absence of the mantle
sublimation region. If the predicted cavity is real, the observed
complex organic molecules must have another origin than grain mantle
sublimation from thermally heated dust. In addition to rising an
important point in itself, the Sch{\"o}ier and J{\o}rgensen et
al. works illustrate the paramount importance of correctly
understanding the physical structure of the source to assess the
chemical structure and all that follows.

For this reason, in the present work, we have re-analyzed the
available data on IRAS16293 {\it from scratch}, considering, in
addition, the most recent evaluation of the distance to this source
by \citet{Loi08} (120 pc instead of 160 pc, as assumed in Ceccarelli
et al., Sch{\"o}ier et al. and J{\o}rgensen et al. works). This new
analysis is necessary and timely because of two important
observational projects having IRAS16293 as a target: a) the unbiased
spectral survey in the 3, 2, 1 and 0.8 mm bands just obtained at the
IRAM and JCMT telescopes \citep[``The IRAS16293-2422 Millimeter And
Sub-millimeter Spectral Survey''\footnote{\it
  http://www-laog.obs.ujf-grenoble.fr/heberges/timasss/}; ][in prep.]{Caux2010in_prep}, and b) the unbiased spectral survey between 500 and 2000
GHz which will be obtained shortly with the heterodyne instrument HIFI aboard
the Herschel Space Observatory (HSO) launched in May 2009 (the
Herschel Guaranteed  Time Key Program CHESS---\emph{Chemical Herschel Surveys
of Star Forming Regions}\footnote{\it
  http://www-laog.obs.ujf-grenoble.fr/heberges/chess/}).  The two
projects, involving large international teams, will provide an
accurate census of the molecular inventory of IRAS16293, the largest
ever obtained in a solar type protostar. To convert the observations
into an accurate chemical composition across the IRAS16293 envelope,
the dust and gas density and temperature profiles have to be
determined accurately first. Deriving these profiles is the goal of the
present article.

We conclude this section by addressing the problem of the binarity of
IRAS16293 and how it fits with the analysis we present here. As soon
as interferometric observations became available it was realized that
IRAS16293 is indeed a proto-binary system \citep{Woo89,Mun92},
composed by two sources: A (the south source) and B (the north
source) separated by $\sim$4$''$, i.e. $\sim$500 AU at 120 pc. While source B is the brightest in the continuum, source A is
often, but not always, the brightest in the molecular emission
\citep[e.g.][]{Cha05}. The most recent observations show that
IRAS16293 is indeed a triple system, with the source A composed by two
objects, A1 and A2, of 0.5 and 1.5 M$_\odot$ respectively \citep{Loinard2009}. While it is clear that the multiple nature of IRAS16293
cannot be neglected in general, the two projects mentioned above 
involve observations with single-dish telescopes, so that much of
the structure on small scales is smeared out in these
observations. Specifically, the molecular line emission will be
dominated by the cold envelope, which fills up the telescope beam, and
by any warm component at the interior of the envelope. The major goal
here is to give a reliable estimate of the envelope temperature and
density profiles of both the gas and dust components, up to the scales
where the approximation of a spherical symmetry is valid. What are
these scales will be discussed later on, based on the available
observations.

The article is organized as follows. Section \ref{sec:dust} discusses
the derivation of the dust density and temperature distribution, based
on the analysis of all available continuum data. Section \ref{sec:gas}
describes the derivation of the gas temperature profile, with the help
of ISO data to constrain the abundance of a major gas
coolant. Finally, Section \ref{sec:conclusions} discusses and
summarizes the results of the presented study.


\section{Dust temperature and density profiles}\label{sec:dust}

\subsection{The data set}\label{sec:dust-data-set}
The present analysis is based on the continuum emission from the
envelope that forms/surrounds the protostar IRAS16293. Three types of
observations are considered: maps of the emission, spectral energy
distribution (SED) and interferometric observations at 1 and 3 mm. All
data used have been retrieved from archives, except of the map at 350 $\mu$m
and the 1 and 3 mm interferometric data that we obtained in dedicated
observations. Below we briefly describe the data used.\\

\noindent
\emph{i) Continuum emission profiles}

We used the maps of the dust continuum emission at 350~$\mu$m
(obtained at the Caltech Submillimeter Observatory; CSO), and 450 and
850 $\mu$m (obtained at the James Clerk Maxwell Telescope; JCMT).

The 450 and 850 $\mu$m maps have been retrieved from the JCMT archive
(website). The beam sizes are 7.5$''$ and 14.8$''$ at 450 and 850 $\mu$m
respectively. Based on the many previous JCMT published observations,
the calibration uncertainty and noise levels are $\lesssim$ 10\% and
0.04 Jy beam$^{-1}$ at 850 $\mu$m and $\lesssim$ 30\% and 0.3 Jy
beam$^{-1}$ at 450 $\mu$m, respectively.

Observations of the 350 $\mu$m continuum emission toward IRAS 16293 reported here were carried out in 2003 February using the SHARC~II facility bolometer camera of the Caltech Submillimeter Observatory (CSO) on Mauna Kea in Hawaii \citep{Dowell03}. SHARC~II is a $12 \times 32$ pixel filled array with a field of view of $2.6 \times 1.0^{\prime}$. The data were taken during excellent submillimeter weather conditions (a 225~GHz zenith opacity of ~0.04, corresponding to less than 1 mm of precipitable water). The observations were carried out using the ``box-scan'' scanning mode (see http://www.submm.caltech.edu/sharc/). Five 10 min scans were reduced together using the CRUSH software package \citep{Kovacs08} to produce the final calibrated image. Telescope pointing was checked on Juno, Vesta, or Mars, immediately before or after the science observations and the measured offsets were applied during data reduction. The data were taken before the CSO Dish Surface Optimization system (DSOS) become operational. The shape of the telescope beam was determined from the pointing images of Vesta and Mars. It contains a diffraction limited main beam with a FWHM diameter of 9$^{\prime\prime}$ and an error beam with a FWHM diameter of 22$^{\prime\prime}$, with relative peak intensities of 0.8 and 0.2, respectively. This size of the error beam is consistent with earlier 350~$\mu$m measurements using the SHARC I camera \citep{Hunter97}.

\begin{figure*}
\includegraphics[width=14cm,angle=0]{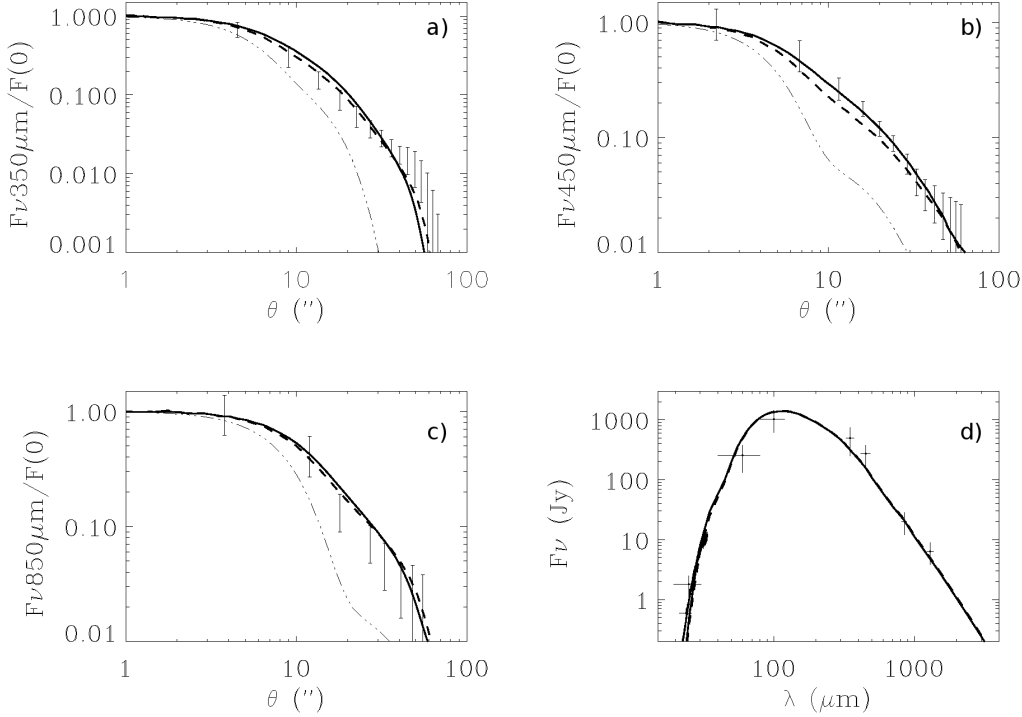}
\caption{Continuum emission profiles at 850 $\mu$m (left bottom
  panel), 450 $\mu$m (right upper panel) and 350 $\mu$m (left upper
  panel), plus the SED (right bottom panel). The curves show the emission predicted by two
  models with different density profiles (see text for more details):
  the solid thick line is the best fit of the maps and SED data
  obtained by a density profile with a 2-indexes power law with
  $\alpha_{out}$=2 and $\alpha_{in}$=1.5 in the outer and inner
  envelope respectively (Shu-like density distribution); the dashed
  line is the best fit obtained by a single power-law index,
  $\alpha$=1.8. The dotted-dashed lines show the telescope beam
  profiles adopted.}
\label{fig:fits}
\end{figure*}

We obtained continuum emission profiles as function of the distance
from the center of the envelope, by averaging the map emission over
annuli at the same distance. In each case we adopted a radial sampling corresponding to the half of the resolution of the instrument. The uncertainties of continuum emission
profiles were evaluated taking into account the calibration
uncertainty, noise levels and the non sphericity of the source. The
resulting profiles, normalized to the peak emission, are shown in
Fig. \ref{fig:fits}. Note that the 450 and 850 $\mu$m profiles are
identical to those reported by \citet{Sch02}.
\\

\noindent
\emph{ii) Spectral Energy Distribution}

Table \ref{SED_tab} reports the SED obtained considering all the data available in the
literature (plus the 350 $\mu$m point obtained by us; see above). The
millimeter and sub-millimeter data points have been obtained by
integrating the maps over the entire envelope. The integrated flux at
1.3 mm was that quoted in \citet{Sar96}. We retrieved the IRAS fluxes
from the IRAS Point Source Catalog v2.1
(\textit{http://irsa.ipac.caltech.edu/cgi-bin/Gator/nph-dd?catalog=iraspsc}).

\begin{table}[bt] \centering
\begin{tabular}{|lllll|} \hline 
$\lambda$   & F$_\nu$\tablefootmark{a} & $\Delta$F$_\nu$\tablefootmark{b} &  $\theta_{mb}$\tablefootmark{c}  & Ref. \\
($\mu$m)    & (Jy)    &         (Jy)    &     (``)        & \\ \hline
23.7        &  0.6    & 0.1             &   6.0           &  2 \\          
25          &  1.8    & 0.7             &   80.0          &  3 \\          
60          &  255.   & 122.            &  160.0          &  3 \\          
100         &  1032.  & 412.            &  237.0          &  3  \\           
350         &  500.   & 250.            &  9.0  &  1  \\              
450         &  270.   & 108.            &  7.8  &  1   \\               
850         &  20.2   & 8.              &  14.5 &  1   \\            
1300        &  6.4    & 2.6             &  22.0 &  4  \\ \hline         
\end{tabular}
\caption[]{The Spectral Energy Distribution of IRAS 16293-2422.\\
\tablefoottext{a}{Integrated flux in Jy}.
\tablefoottext{b}{Uncertainties in Jy considering the calibration uncertainty, the noise levels and the uncertainty on the source size.} 
\tablefoottext{c}{Main beam of the instrument in arcsec.}
} 
\tablebib{(1) This paper; (2) Spitzer catalog; (3) IRAS Point Source Catalog v2.1; (4) \citet{Sar96}.}
\label{SED_tab}
\end{table}

IRAS 16293 was observed with the InfraRed Spectrograph (IRS)
installed aboard the {\it Spitzer Space Telescope} as part of the
``From Molecular Cores to Planet Forming DIsks'' \citep{Evans2003,Evans2005} Legacy Program
(AOR: 11826944, PI: Neal Evans). We use the observations obtained with
the Long-High (LH) module (20-37 $\mu$m, R $=$ 600) on 
2004 July 29, in staring mode. The data reduction was performed using the
c2d 
pipeline S15.3.0 \citep{Lahuis2006} with the pre-reduced (BCD)
data. Since the MIPS map at 24 $\mu$m shows that the IRAS16293
emitting region ($\sim$ 30-40$''$) is larger than the LH module field of
view (11.1$''$ $\times$ 22.3$''$), we adopted the full aperture extraction
method in the pipeline. In addition, we corrected the derived flux
level for the missing flux by comparing the IRS spectrum integrated on
the MIPS bandwith and the integrated MIPS flux. Note that this method
assumes a similar distribution of the source emission in the whole
IRS wavelength interval, $\sim$ 21-37 $\mu$m. The correction factor
derived by this method amounts to 22 $\%$ in the flux.\\

\noindent
\emph{iii) Interferometric continuum data}

Observations of the 3 and 1.3 mm continuum were obtained with the
IRAM Plateau de Bure Interferometer (PdBI) in 2004 and are described
in detail in \citet{Bot04}. They were obtained in B and C
configurations of the PdBI array, resulting in a spatial resolution of about
0.8$''$.

\subsection{Adopted model}
We obtained the best fit to the continuum data, described in the
previous section, by using the 1D radiative transfer code DUSTY
\citep{Ive97}, which has been extensively used in similar works
\citep[including][]{Jor02,Sch02,Sch04,Jor05}. Briefly, giving
as input the temperature of the central object and a dust density
profile, DUSTY self-consistently computes the dust temperature profile
and the dust emission. A comparison between the computed 350, 450,
850 $\mu$m brightness profiles (namely the brightness versus the
distance from the center of the envelope) and SED with the observations
(described in the previous section) allows to
constrain the density profile and, consequently, the temperature
profile of the envelope.  To be compared with the observations, the
theoretical emission is convolved with the beam pattern of the
telescope. Following the recommendations for the relevant telescope,
the beam is assumed to be a combination of Gaussian curves: at 850
$\mu$m, we use HPBWs of 14.5$''$, 60$''$, and 120$''$, with amplitudes of
0.976, 0.022, and 0.002 respectively; at 450 $\mu$m, the HPBWs are
8$''$, 30$''$, and 120$''$ with amplitude ratios of 0.934, 0.06, and
0.006, respectively \citep{San01}; at 350 $\mu$m, we use HPBWs of
9$''$ and 22$''$, with amplitude ratios of 0.8, 0.2, respectively
(Section 2.1).

In this work, we consider two cases for the density distribution. In
the first one, we assum a broken power-law density profile as
in the \citet{Shu77} solution: 
\begin{equation}\label{shu_inf}
n(r)=n(r_0)  \left( \frac{r_0}{r} \right)^{1.5}~~~~~r\leq r_{inf}
\end{equation}
\begin{equation}\label{shu_stat}
n(r)=n(r_0)  \left( \frac{r_0}{r} \right)^{2}~~~~~r\geq r_{inf}
\end{equation}

where $r_{inf}$ is the radius of the collapsing envelope (at larger
radii the envelope is static). In the density profile, it represents
the radius at which the change of index occurs and it is a free
parameter. Note that the complete \citet{Shu77} solution also contains a transition part at the interface of the collapsing and static regions, just inside $r_{inf}$. In this region the slope gradually changes from r$^{-1}$ to the limiting value of r$^{-1.5}$. A posteriori, the effective part of the envelope in r$^{-1}$ is relatively small and located in the inner region ($\lesssim$ 1500 AU, equivalent to $\sim$ 10''). Since the continuum observations, with resolutions of $\sim$ 8-18$``$, are not sensitive enough, we adopted the simplified structure described by the Eq. \ref{shu_inf},\ref{shu_stat}.
 In the second case, we considered a single power-law
density profile, where the index $\alpha$ is a free parameter:
\begin{equation}
  n(r)=n(r_0)  \left( \frac{r_0}{r} \right)^\alpha
\end{equation}
In both cases, $n(r_0)$ is the density at $r_0$, and the envelope starts
at a radius $r_{in}$ and extends up to $r_{out}$. In total, both
models have four free parameters determined by the best fit with the
observational data: $r_{inf}$ or $\alpha$, $n(r_0)$, $r_{out}$ and
$r_{in}$. Finally, DUSTY requires the temperature of the central
source, T$_*$, here assumed to be 5000 K. Note that we verified that
the choice of this parameter does not influence the results, as
already noticed by other authors \citep[e.g.][]{Jor02}. In
practice, the DUSTY input parameters are the infall radius $r_{inf}$
or the power-law index $\alpha$, the optical thickness at 100 $\mu$m,
$\tau_{100}$, the ratio between the inner and outer radius, Y
(=$r_{out}$/$r_{in}$) and the temperature at the inner radius
T$_{in}$\footnote{The temperature at the inner radius T$_{in}$ in fact
  defines the radius at which the integration starts. The DUSTY codes
  requires the temperature rather than the radius because it is based
  on a scale-free algorithm.}. The optical thickness is, in turn,
proportional to $n(r_0)$ and $r_{out}$.  In both models, we obtain a
lower limit to T$_{in}$ of 300 K, any larger value giving similar
results.

In addition to the above parameters, the opacity of the dust as
function of the wavelength is a hidden parameter of DUSTY. Following
numerous previous studies \citep{Van99,Eva01,Shi02,You03,Sch02}, we
adopted the dust opacity calculated by \citet{Oss94}, specifically
their OH5 dust model, which refers to grains coated by ice. Again, the
basic result, though, does not substantially depend on the choice of
the dust opacity model.

As explained in \citet{Ive97}, DUSTY gives scale-free
results, so that the source bolometric luminosity L$_{bol}$ and the
distance are required to compare the DUSTY output with actual
observations. We assumed here the lastest estimate of the distance to
the $\rho$
Ophiuchus cloud, namely 120 pc (see Introduction), and we derived the
bolometric luminosity by integrating the observed emission over the
full spectrum, optimizing the resulting $\chi^2$ (see below).

We run grids of models for both cases described above.
The summary of the covered parameter space is reported in Table
\ref{DUSTY_input}.
\begin{table}[bt] \centering
\begin{tabular}{|lll|} \hline \hline 
Parameter   & Range   & Step \\ \hline
$\alpha$    & 0.2--2.5  &  0.1    \\ 
Y$^{a}$         & 50--2000   & 10  \\
r$_{inf}$    &  5--200        & 2    \\ 
$\tau_{100}$ & 0.1--10.  & 0.1 \\
T$_{in}$     & 300 K    & Fixed \\
T$_*$       & 5000 K    & Fixed \\ \hline 
\end{tabular}
\caption[]{Range and step of the DUSTY input parameters.\\\hspace{\linewidth}
$^{a}$ Y = r$_{out}$/r$_{in}$} \label{DUSTY_input}
\end{table}
The best fit model has been found minimizing the $\chi^2$ with an
iterated two-steps procedure \citep[see also][]{Crim2009}.  First,
we use the observed brightness profiles at 350, 450 and 850 $\mu$m to
constrain Y and $\alpha$ (or r$_{inf}$ in the case of the Shu-like
model), assuming a value for $\tau_{100}$. Second, we constrain the
optical thickness $\tau_{100}$ by comparing the computed and observed
SED, assuming the $\alpha$ (or r$_{inf}$) and Y of the previous
step. The new $\tau_{100}$ is used for a new iteration and
so on. In practice, the iteration converges in two steps. This is
because the normalized brightness profiles depend very weakly on
$\tau_{100}$, while they very much depend on the assumed size of the envelope
and on the slope of the density profile \citep[see also the discussion in][]{Jor02,Crim2009}. In contrast, the
dust optical thickness depends mostly on the absolute density of the
envelope.  Note that we constrain also the bolometric luminosity based
on the best fit of the SED \footnote{Note that the dust optical
  thickness $\tau_{100}$ affects the shape of the SED, so in general
  it enters in the determination of the bolometric luminosity.}.

\subsection{Results}

\noindent
\emph{a) Brightness profiles and SED analysis}

Table \ref{best_fit_param} lists the best fit parameters for the two
models described in the previous section as well as the parameters
obtained by \citet{Sch02} for comparison.  Figure \ref{fig:fits} shows
the solutions compared to the observed data (maps and SED).

\begin{figure*} \centering
\includegraphics[width=10cm,angle=0]{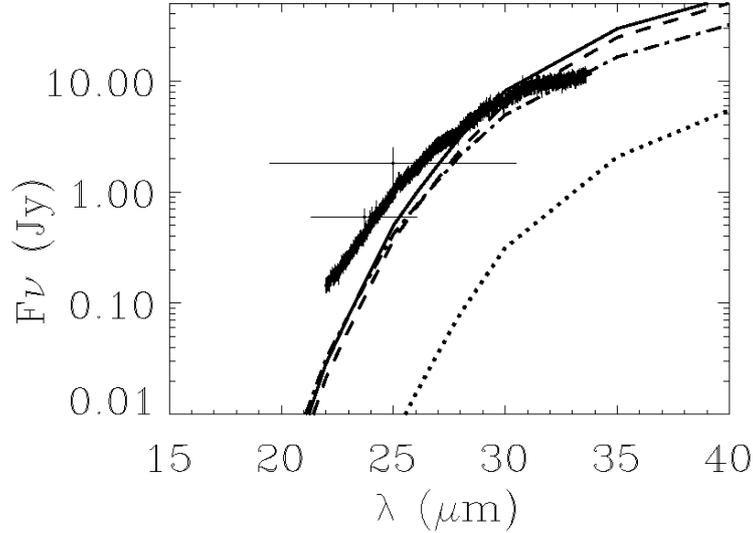}
\caption{MIR Spitzer and IRAS fluxes as a function of the wavelength. The two crosses represent the integrated fluxes at 23.7 and 25 $\mu$m obtained with MIPS aboard Spitzer and CPC aboard IRAS, respectively. The solid thick line is the IRS SPITZER spectrum observed between $\sim$ 21--37 $\mu$m (see text for details). The solid thin line is the best fit of the maps and SED data
  given by the model of envelope with a density profile with a
  broken power law with
  $\alpha_{out}$=2 and $\alpha_{in}$=1.5 in the outer and inner
  envelope respectively (Shu-like density distribution). The dashed
  line is the best fit given by the model of envelope with a single power-law index,
  $\alpha$=1.8. The dotted-dashed line and the dotted lines show the emission resulting from the models obtained by \citet{Jor05} and \citet{Sch02}, respectively.}
\label{SED_spitzer}
\end{figure*}

\begin{figure*} \centering
  \rotatebox{0}{\includegraphics[width=9cm,angle=90]{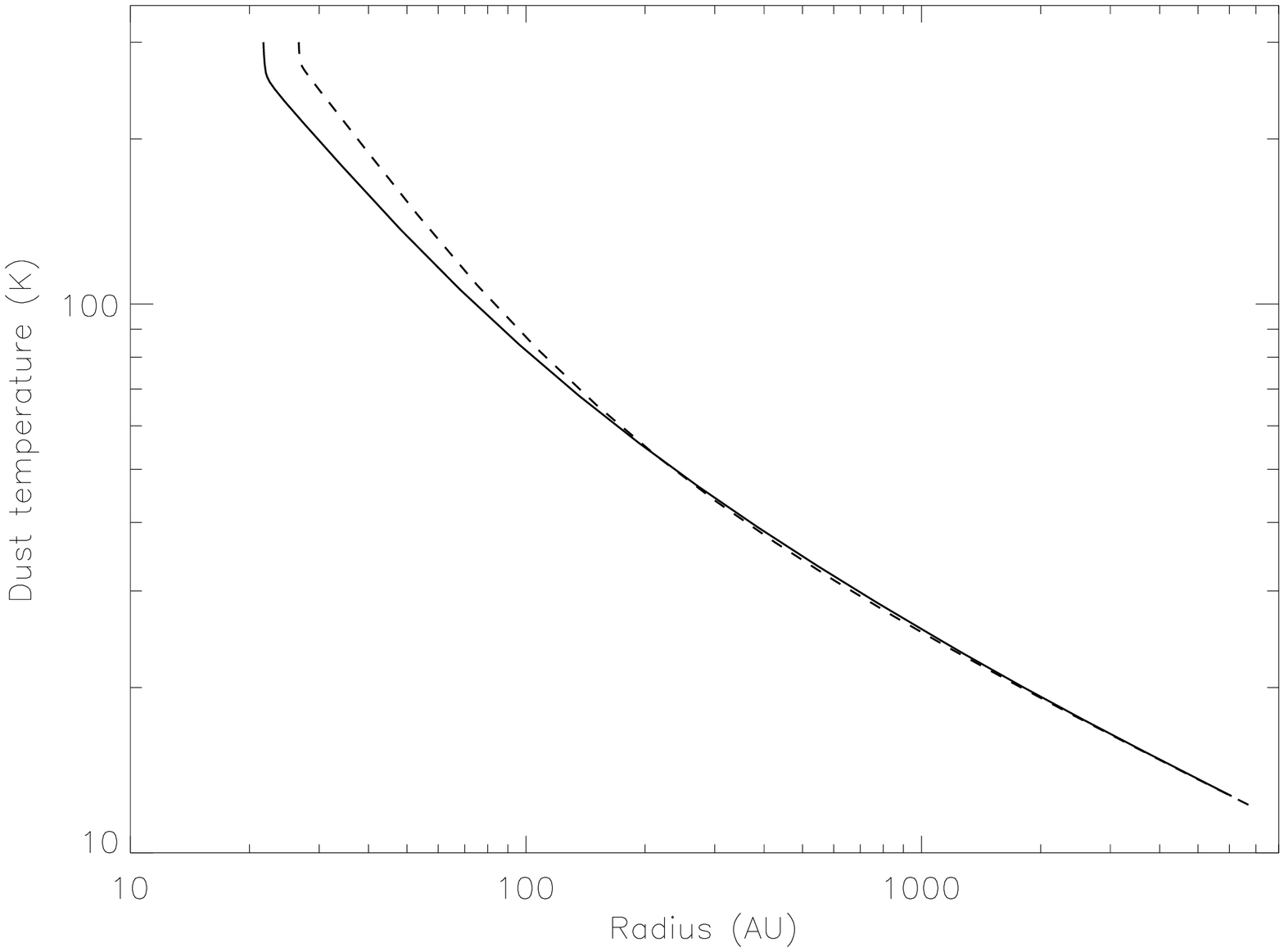}}
  \rotatebox{0}{\includegraphics[width=9cm,angle=90]{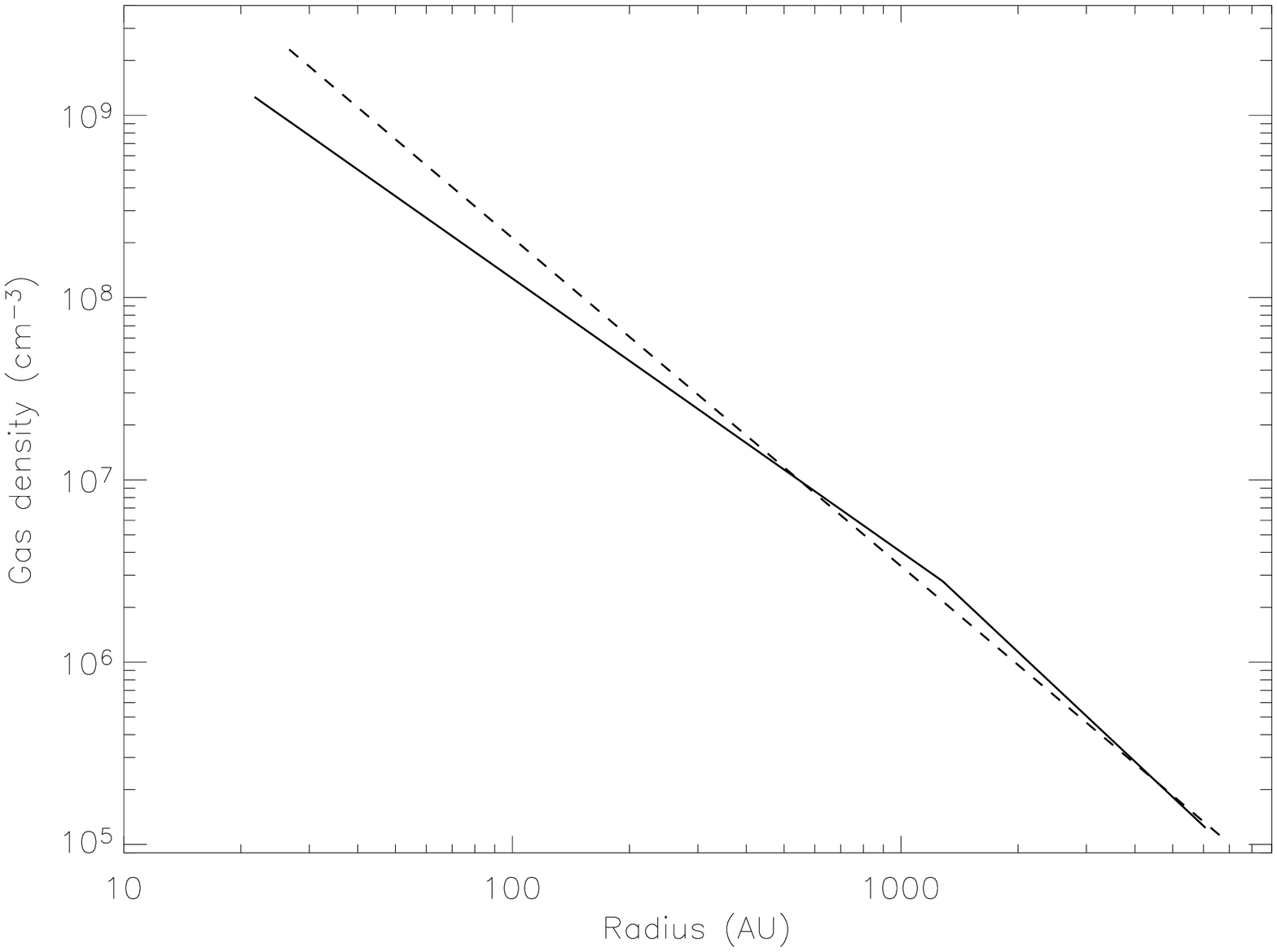}}
  \caption{Dust temperature (upper panel) and H$_{2}$ density (lower
    panel) profiles from the best fit obtained in the two cases Shu-like
    (solid line) and single power-law density distribution (dashed
    line). }
  \label{T_n}
\end{figure*}

\begin{table*}
  \begin{tabular}[bt]{lccc}
    \hline
    Quantity          & Shu-like model & Power-law model & \citet{Sch02} \\ \hline
    \multicolumn{4}{c}{$\mathrm{Best~fit~parameters}$} \\
    L$_*$ (L$_\odot$)  & 22             & 22   & 27 \\
    D (pc)             & 120            & 120  & 160\\
    $\tau_{100}$       & 2.0            & 3.0  & 4.5\\
    $r_{inf}$ (AU)     & 1280           &  &  \\
    $\alpha$          &                & 1.8 & 1.7\\
    Y                 & 280            & 260 & 250\\    \hline
    \multicolumn{4}{c}{$\mathrm{Physical~quantities}$} \\
    \hline
    $r_{in}$ (AU)     & 22             & 27 & 32\\
    $r_{out}$ (AU)    & 6100           & 6900 & 8000 \\
    r(T$_{dust}$=100K) (AU) & 76       & 85 & \\
    n(T$_{dust}$=100K) (cm$^{-3}$) & $2\times10^{8}$ & $3\times10^{8}$ & \\
    M$_{env}$ (M$_\odot$) & 1.9         & 2.1 & 5.4 \\
 \hline
  \end{tabular}
  \caption{Summary of the dust radiative transfer analysis. The upper 
    half of the table lists the best fit parameters, the lower half reports 
    some relevant physical quantities. The last column reports the results 
    of the \citet{Sch02} analysis for comparison. \label{best_fit_param}}
\end{table*}
The results of this part of the analysis are:
\begin{itemize}
\item The bolometric luminosity, obtained minimizing the $\chi^2$, is
  22 L$_\odot$ rather then 14 L$_\odot$, the value obtained by simply
  scaling the bolometric luminosity adopted by \citet{Cec00a} and
  \citet{Sch02} to a distance of 120 pc instead of 160 pc.

\item The single power-law and Shu-like density distributions can both
  reproduce the observations, the maps and the SED data, included the
  IR part of the spectrum (Figure \ref{fig:fits}). They give similar best fit $\chi^2$ values,
  although the Shu-like distribution fits better the 450 $\mu$m
  profile. Also the derived physical parameters are substantially
  similar for the two models.

\item The observed data, including the Spitzer data between 20 and 40
  $\mu$m, do not necessitate the presence of a large cavity
  \citep{Jor05}. Our solution with a larger luminosity (22 instead of
  14 L$_\odot$) can reproduce as well the MIR Spitzer observations. Figure \ref{SED_spitzer} shows the MIR Spitzer and IRAS observations against the emission predicted by our models and these by \citet{Jor05} and \citet{Sch02}.
\end{itemize}
Figure \ref{T_n} shows the derived dust temperature and density
profiles. The single power-law index density distribution predicts a
slightly warmer and denser region at radii lower than about 100 AU,
but the differences are relatively small (see also
Table \ref{best_fit_param}).
\\

\noindent
\emph{b) Interferometric data analysis}

The previous analysis considers single dish data, which at best have a
spatial resolution of 8$''$, equivalent to a radius of $\sim 1000$ AU.
In order to constrain better the inner region, we used 1 and 3 mm
continuum interferometric observations obtained with the Plateau de
Bure interferometer (PdBI) described in the previous section. In order to
compare the model predictions with the observations, we have produced
synthetic maps for each model, in which we added the predicted
emission from the envelope plus the emission of the two Gaussian-like
sources A and B. The parameters used to represent the emission of the sources A and B are extracted from the interferometric maps at 1.3 and 3 mm and reported in Table \ref{gauss_param}. Then we used the UV$\_$FMODEL task in Gildas\footnote{http://iram.fr/IRAMFR/GILDAS/doc/html/mis-html/node13.html} to compute visibility tables with the same $\emph{uv}$ plane coverage
than the actual observations. 

\begin{table}[bt] \centering
\begin{tabular}{|l|ll|ll|} \hline 
            &   &1.3mm        &  &3mm           \\ \hline
Source   & F$_{\nu}$(Jy)   & FWHM('') &  F$_{\nu}$(Jy) & FWHM('') \\ \hline
A        & 0.49    & 1.18     & 0.16   & 1.62     \\ 
B        & 0.97    & 0.86     & 0.25   & 0.78      \\ \hline 
\end{tabular}
\caption{Integrated flux and Full Width at Half Maximum (FWHM) of the gaussians used to model the emission of the sources A and B at 1.3 and 3 mm.} \label{gauss_param}
\end{table}

\begin{figure*} \centering
 \includegraphics[width=15cm,angle=0]{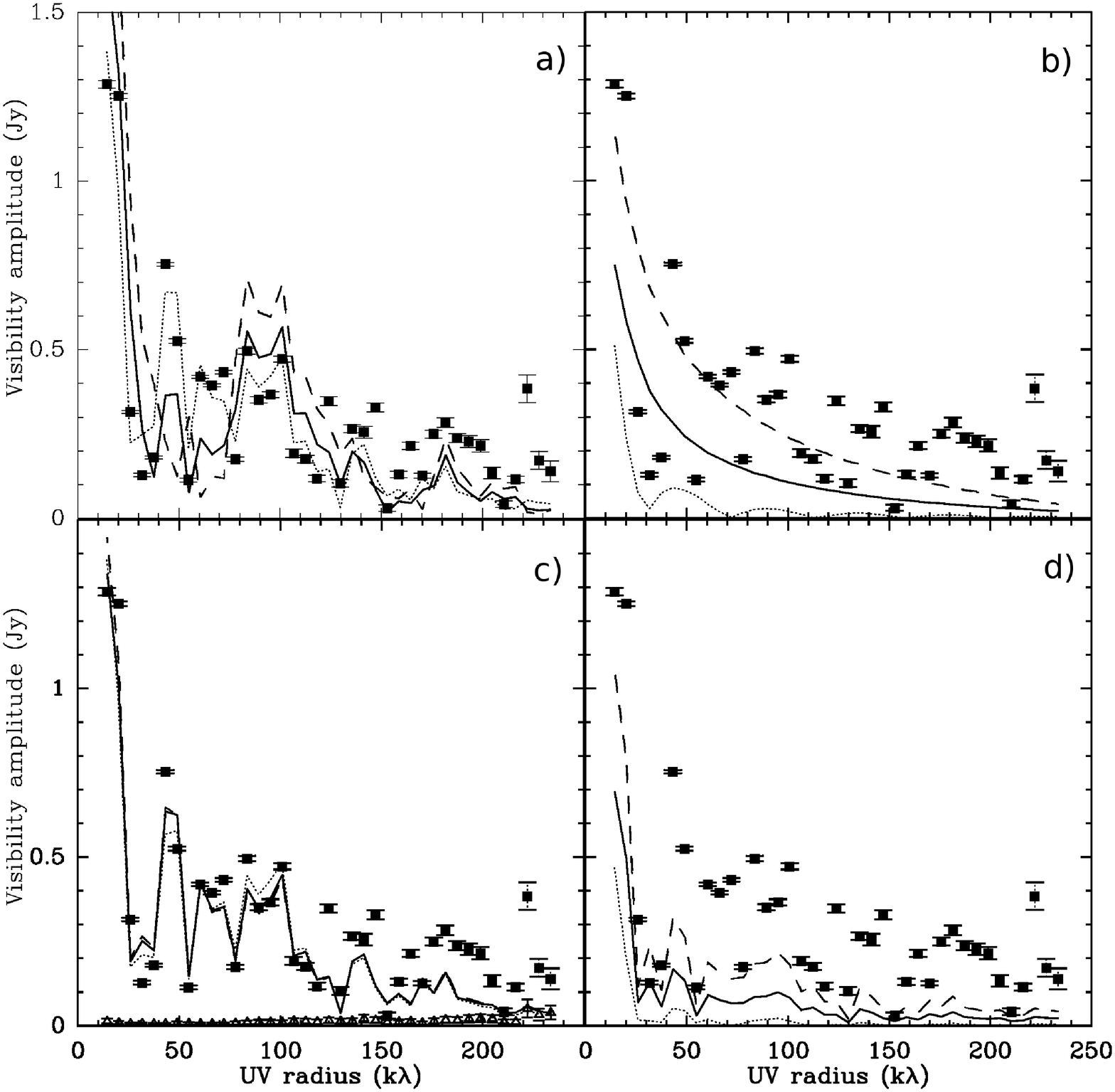}
  \caption{Visibility amplitudes as a function of the $\emph{uv}$ radius of the
    1.3 mm PdBI observations (squares) against the predictions
    obtained with the best fit models of a Shu-like (solid lines) and
    single power-law (dashed lines) density distribution. The model of
    \citet{Sch04}, with the central cavity, is shown for a comparison (dotted lines). 
Panels (a) and (b) show the case where the envelope is centered on the mid-way point between the sources A and B. Panels (c) and (d) show the case where the envelope is centered on the source B. Panels (a) and (c) show the visibility amplitudes modeled with the synthetic maps containing the contribution of the envelope and the sources A and B. Panels (b) and (d) show the visibility amplitudes modeled with the synthetic maps containing the contribution of the envelope only. Finally, the zero-signal expectation values are represented by the empty triangles on the c panel.}
  \label{pdb}
\end{figure*}

Table \ref{X2_interfero} reports the $\chi^2_{red}$ values
  obtained using the observed and modeled visibility amplitudes at 1.3
  mm. The comparison is done for each of the three models of the envelope, i.e. the single power-law density distribution, the Shu-like  density distribution, and the central cavity of \citet{Sch04}. We tested cases in which the envelope is centered on one of the two sources or on the mid-way point between the two sources.
The $\chi^2_{red}$ values are computed over more than 10$^4$ points. The 2D representation of the visibility amplitudes observed and modeled in $\emph{uv}$ plan is difficult to read. Therefore to illustrate the comparison between the visibility amplitudes observed and modeled, we averaged the visibility amplitudes over the same $\emph{uv}$ radius and plotted them in Figure \ref{pdb}. Note that the error bars shown in Fig. \ref{pdb} represent only the measurement errors. In fact, the standard deviations resulting from the radial average are meaningless in this case because of the non-sphericial geometry (caused by the presence of the sources A and B). Figure \ref{pdb} only aims to illustrate the comparison between the observations and models, our conclusions will be based on the $\chi^2_{red}$ in Table \ref{X2_interfero}. The computations were
  done considering all the visibility amplitudes obtained in the  $\emph{uv}$ plane before the azimuthal average.

The interferometric data
are dominated by the two components of the IRAS16293 binary system. However, regarding
the envelope contribution---which is relevant for the present work---in
the case where the envelope is centered on the mid-way point between the sources A and
B, the visibilities at $\emph{uv}$ radii lower than about 80 k$\lambda$ are not
well reproduced by the single power-law density profiles. Our solution
with the Shu-like density profile and the solution with a cavity
(suggested by \citealt{Sch04}) give a much better agreement. However, when envelope is centered on B or A
all three models give similar results (Fig. \ref{pdb}), and fit well
the observed visibilities. Since with the data available in the
literature there is no way to know whether the envelope is centered on
one of the two sources (A or B) or just on the mid-way point between them, either of the
two solutions (centered on a source or on the mid-way point
  between A and B) is equally plausible. In other words, we
  are in a degenerated case dominated by the parameters relative to
  the binarity of the source. The interferometric data can be
  reproduced without a cavity.
Note that the 3 mm PdBI observations do not provide additional information.
We emphasize that the above analysis of the interferometric data is
directly applicable to the OVRO data that led \citet{Sch04} to suggest
the presence of a cavity. The OVRO data probe visibility amplitudes
lower than about 60 k$\lambda$, a range also probed by the PDBI
data. Therefore, as shown above, they can be reproduced by assuming
that the envelope is centered on either A or B source without the
necessity of assuming a cavity.\\
\begin{table*}[h] \centering
\begin{tabular}{|l|c|c|c|} \hline 
$\lambda =$ 1.3 mm & Model centered    & Model centered &  Model centered  \\ 
                   & in between A and B  & on B           &  on A  \\ \hline
Model      & $\chi^2_{red}$ & $\chi^2_{red}$ & $\chi^2_{red}$  \\ \hline
Single power-law    & 18.5   & 10.2  &  8.5  \\ 
Shu-like            & 13.3   & 10.3  &  8.3  \\ 
Central cavity      & 11.0   & 11.3  &  9.9  \\ \hline 
\end{tabular}
\caption{$\chi^2_{red}$ values obtained comparing the visibility
  amplitudes observed and modeled at 1.3 mm. The number of
  degrees of freedom $\nu$ is 10230. The 1$^{st}$ column reports the
  $\chi^2_{red}$ obtained using the synthetic maps with the envelope
  centered on the mid-way point between the sources A and B. The
  2$^{nd}$ and the 3$^{rd}$ columns report the $\chi^2_{red}$ obtained
  using the synthetic maps with the envelope centered on the source B
  and on the source A, respectively. The 1$^{st}$, the 2$^{nd}$, and
  the 3$^{rd}$ lines report the $\chi^2_{red}$ obtained in all the
  cases discussed in the text, using the model of envelope with a single power-law density distribution, a Shu-like  density distribution, and a central cavity of \citet{Sch04}, respectively.}\label{X2_interfero}
\end{table*}

\noindent
\emph{c) Summary of the continuum data analysis}

Both the single dish and interferometric continuum data can be
reproduced by either of the two models we considered: a Shu-like and
single power-law density distribution.  Furthermore, no cavity is
required to explain the data, neither the Spitzer MIR data
\citep{Jor05} nor the interferometric data \citep{Sch04}.

Therefore, in the following we will adopt the Shu-like density profile
(which has a physical interpretation), with no cavity, as our
reference model for the study of the gas temperature and water line
predictions.

\section{Gas temperature profile}\label{sec:gas}

\subsection{Adopted method}
We computed the gas temperature profile using the CHT96 code described
in \citet{Cec96} \citep[see also][]{Cec00b,Mar02,Crim2009}. Briefly, the code computes the gas equilibrium temperature
at each point of the envelope, by equating the heating and cooling
terms at each point of the envelope. Following the method
described in \citet{Cec96}, we considered heating from the gas
compression (due to the collapse), dust-gas collisions and
photo-pumping of H$_2$O and CO molecules by the IR photons emitted by
the warm dust close to the center\footnote{Cosmic rays ionization is a
  minor heating term in the protostellar envelopes.}. The cooling terms are the line
emission from H$_2$O, CO and O. The dust-gas collisions are a source of
heating when the dust temperature is higher than the gas temperature
and a source of cooling elsewhere. 
To compute the cooling from the lines we used the code described in
\citet{Cec96,Cec03} and \citet{Par05}. The same code has been used in
several past studies, whose results have been substantially confirmed
by other groups (e.g. the analysis on IRAS16293-2422 by
\citealt{Sch02}). Briefly, the code is based on the escape probability
formalism in presence of warm dust \citep[see][]{Tak83}, where the
escape probability $\beta$ is computed at each point by integrating
the line and dust absorption over the solid angle $\Omega$ as follows:
\begin{equation} 
  \beta = \frac{k_\mathrm{d}}{k_\mathrm{L} + k_\mathrm{d}} + 
  \frac{k_\mathrm{L}}{(k_\mathrm{L} + k_\mathrm{d})^2} \int d\mu 
  \frac{1-\exp \left[ - \left( k_\mathrm{L} + k_\mathrm{d} \right) 
  \Delta L_\mathrm{th} \right]} {\Delta L_\mathrm{th}} 
\end{equation} 
\noindent  
where $k_\mathrm{L}$ and $k_\mathrm{d}$ are the line and dust
absorption coefficients respectively, and $\Delta L_\mathrm{th}$ is
the line trapping region, given by the following expressions:
\begin{equation} 
  \Delta L_\mathrm{th} = 2 \Delta v_\mathrm{th}  
  \left( \frac{v}{r} \left| 1-\frac{3}{2} \mu^2 \right| \right)^{-1}  
\end{equation} 
\noindent
in the infalling region of the envelope (where $\mathrm{arcos}~
\mu$ is the angle with the radial outward direction) and
in the static region (where $R_\mathrm{env}$ is the envelope radius):
\begin{equation} 
  \Delta L_\mathrm{th} = r \left( 1 - \frac{r}{R_\mathrm{env}} \right) .
\end{equation} 
\noindent 
 In addition, H$_2$O and CO molecules can be pumped by absorption of the NIR photons emitted by
  the innermost warm dust. Since the densities and temperatures of the
  regions of the envelope targeted by this study are not enough to
  populate the levels in the vibrational states, the effect of the NIR
  photons is an extra heating of the gas, as described in
  Ceccarelli et al. (1996). Note that the code takes into
  account the dust with temperatures up to 1500 K.

The code has a number of parameters, that influence the gas equilibrium
temperature. First, the dust temperature, assumed to be the
output of the previous step analysis (\S \ref{sec:dust}). Second,
since the infall heating depends on the velocity gradient across the
envelope and the cooling depends on the gas lines (which can and are
sometimes optically thick), the velocity field across the envelope is
also a parameter of the code. Here we assumed that the infall velocity
field corresponds to the free fall velocity field with a 2 M$_\odot$
source at the center of the envelope \citep{Loinard2009}. Third we
adopted the ``standard'' cosmic rays ionization rate, namely
$3\times10^{-17}$ s$^{-1}$, however, in practcie this parameter is
unimportant. Finally, given the contributions of the H$_2$O, CO and OI
to the gas cooling, their respective abundances are important
parameters of the model. Previous theoretical studies have shown that
the O abundance is constant across the envelope, except in the very
inner regions, where the gas temperature exceeds about 250 K and
endothermic reactions that form OH and H$_2$O become very efficient
\citep{Cec96,Dot97, Dot04}. Unfortunately, this parameter is very
poorly constrained by observations (because of the difficulty of
observing the O fine structure lines and the fact that they are easily
excited in the foreground molecular cloud and its associated PDR 
\citep[see e.g.][]{Lis99,Cau99}. Here we assumed that the atomic O abundance
is equal to $1\times10^{-5}$ and verified a posteriori that the O fine
structure line emission is consistent with the ISO observations. We
note, however, that O is never the dominant coolant except, perhaps,
in a very small region of the envelope \citep{Cec00b,Mar02}. CO, in
contrast, is the main coolant in the outer envelope, but, since
the cooling lines are heavily optically thick, its abundance does not
play an important role (in the regime where it is higher than about
$1\times10^{-6}$ with respect to H$_2$). We, therefore, assumed a
canonical $1\times10^{-4}$ CO abundance across the
envelope.\footnote{Note that it possible that a region where CO
  abundance is lower, because of the freezing onto the dust grains,
  exists. However, the gas cooling is relatively insensitive to this
  lower abundance for the reasons explained in the text.}

Finally, water is an important coolant in the inner region, where the
grain mantles sublimate, injecting into the gas phase large quantities
of water molecules. Again, based on previous studies, we approximated
the H$_2$O abundance with a step function: the abundance is X$_{in}$
in the region where the dust temperature exceeds 100 K, and X$_{out}$
elsewhere. Both X$_{in}$ and X$_{out}$ are found by comparing the
theoretical predictions with the ISO observations (see \S
\ref{sec:gas-results}).  Note that, to solve the water level
population statistical equilibrium equations, we used the collisional
coefficients between H$_2$O and H$_2$ recently computed by
\citet{Fau07}. We assumed that the ortho-to-para H$_2$ ratio is at the
Local Thermal Equilibrium (LTE) in each part of the envelope.
Finally, we assumed a H$_2$O ortho-to-para ratio equal to 3.

\subsection{Results}\label{sec:gas-results}

\begin{figure} \centering
  \rotatebox{0}{\includegraphics[width=9cm]{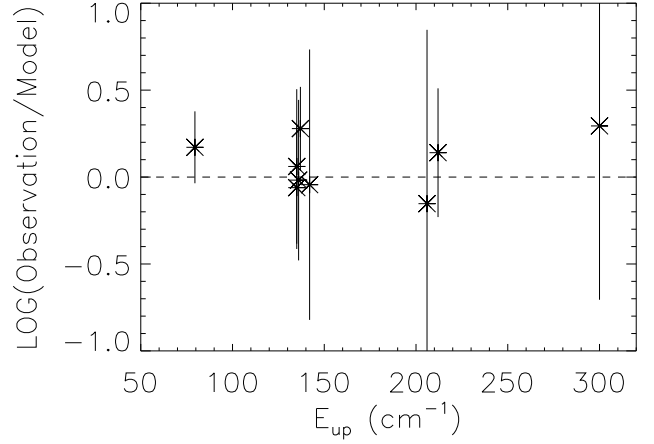}}
  \caption{Ratio of the observed over predicted H$_2$O line flux
    versus the upper level energy in cm$^{-1}$. In this computations,
    we used the dust temperature and density profiles of the Shu-like
    model of Table \ref{best_fit_param} and an H$_2$O abundance equal
    to $5\times10^{-8}$ in the inner envelope and $2.5\times10^{-7}$
    in the outer envelope (see text). }
  \label{h2o-obsmod}
\end{figure}

In order to constrain the water abundance, necessary to predict the
gas temperature profile, we used the ISO observations of the water
lines towards IRAS16293 \citep{Cec98,Cec00b} and compared
with the model predictions obtained with different H$_2$O abundance
profiles.  For that, we run a grid of models with X$_{in}$ and X$_{out}$ varying
between $1\times10^{-8}$ and $1\times10^{-5}$.
\begin{figure} \centering
  \rotatebox{0}{\includegraphics[width=9cm]{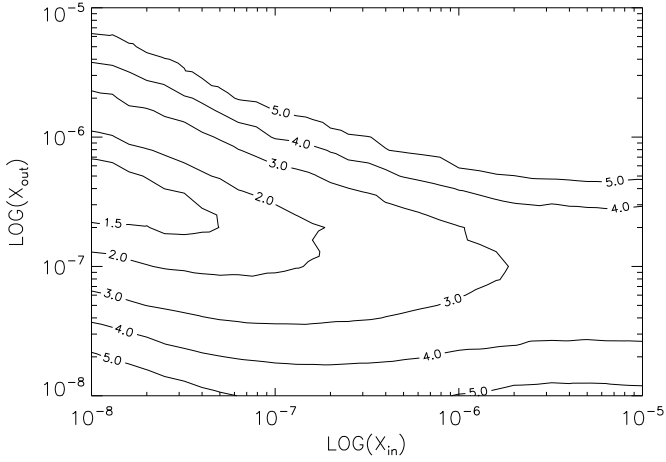}}
  \caption{Reduced-$\chi^2$ contours obtained comparing the model
    predictions and the observations towards IRAS16293, as function of
    the inner X$_{in}$ and outer X$_{out}$ H$_2$O abundance. In this
    computations, we used the dust temperature and density profiles of
    the Shu-like model of Table \ref{best_fit_param}. }
  \label{h2o-chi2}
\end{figure}
Figure \ref{h2o-chi2} shows the resulting $\chi^2$ as function of the
inner and outer H$_2$O abundance, in the case of the Shu-like
structure of Table \ref{best_fit_param}. The H$_2$O abundance in the
outer envelope is constrained between 0.7--20 $\times10^{-7}$ at 3
$\sigma$, with the best fit solution at $2.5\times10^{-7}$ (reduced
$\chi^2$=1.3). The inner H$_2$O abundance is even less constrained,
lower than about $2\times10^{-6}$ at 3 $\sigma$ confidence level, and
$0.5\times10^{-7}$ at 1.5 $\sigma$ confidence level, i.e. slightly lower than the
inner abundance, which is in contradiction with the hypothesis that
ice mantles sublimate injecting water in the gas phase. However, one
has to consider that the inner H$_2$O abundance is very poorly
constrained by the ISO observed lines, which are very optically thick
and not high enough in energy (as it is shown by the predictions of
Table \ref{predictions_PACS}). We, therefore, only consider significant
the 3$\sigma$ level limit to the water inner abundance, warning that
even that may be questionable for the same reasons (see also \S
\ref{sec:conclusions}). The ratio between the observed by ISO and
predicted lines fluxes as function of the upper level energy is shown
in Fig. \ref{h2o-obsmod}.

The results for the single power-law density profile are similar, with
the upper limit on the inner H$_2$O abundance a factor ten lower
(because of the higher density in the inner part predicted by this
model).

\begin{figure*} \centering
  \rotatebox{0}{\includegraphics[width=9cm,angle=90]{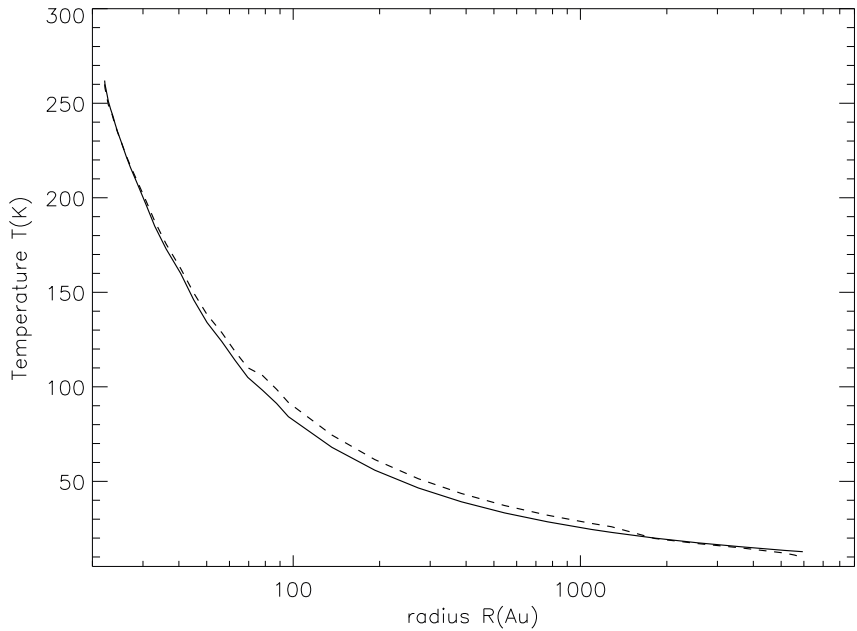}}
  \caption {Dust (solid line) and gas (dashed line) temperature
  profiles of the best-fit obtained assuming a Shu-like density
  distribution (see text).\label{Tgas-profile}}
  
\end{figure*}
\begin{figure*} \centering
\includegraphics[width=9cm,angle=90]{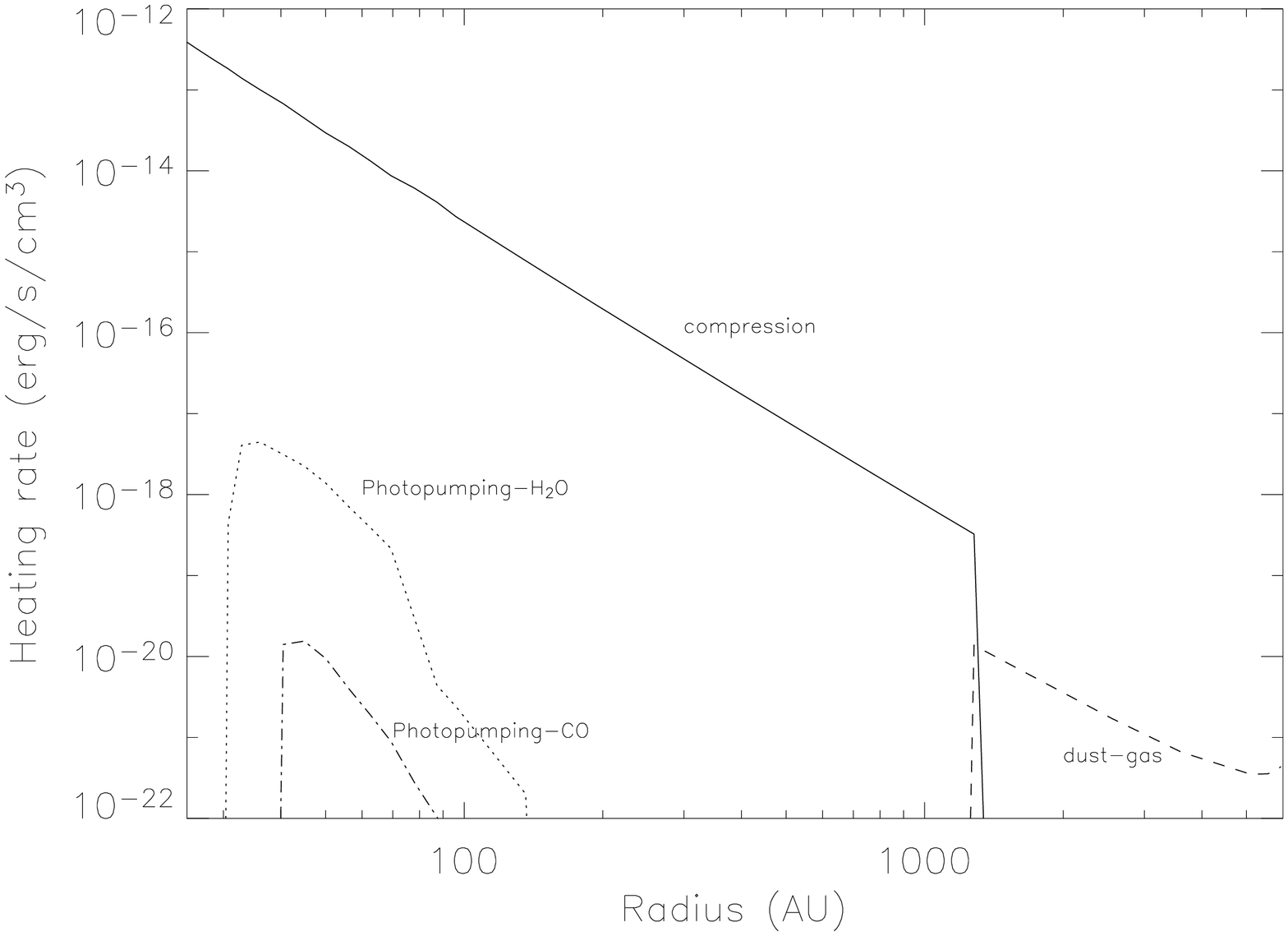}
\includegraphics[width=9cm,angle=90]{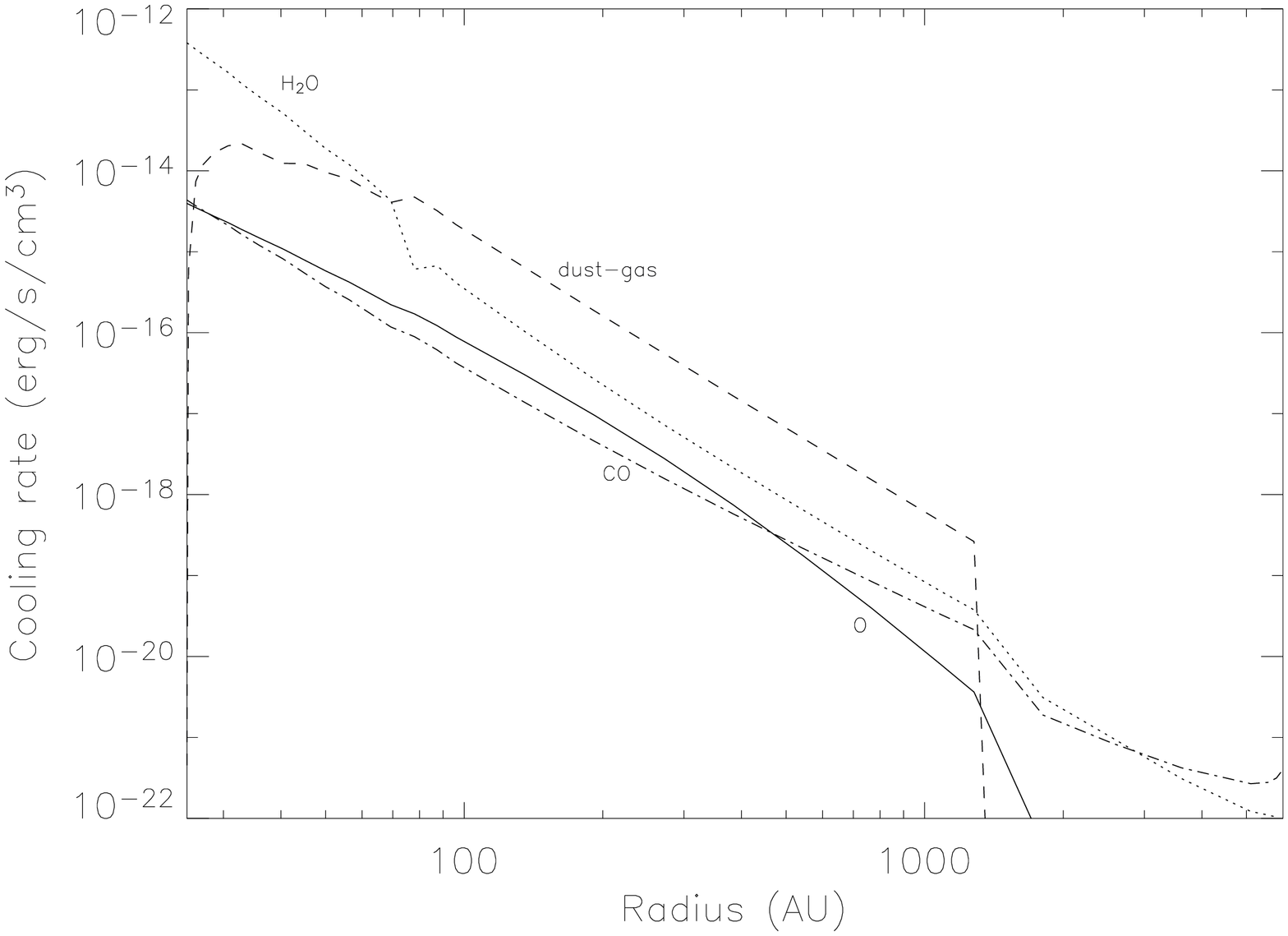}
\caption {Heating (top panel) and cooling (bottom panel) rates as
  function of the radius, computed assuming that the inner H$_2$O
  abundance is equal to $2\times10^{-6}$ and the outer H$_2$O
  abundance is $2.5\times10^{-7}$, for the Shu-like best fit model. 
  The solid line shows the compression 
  heating (top panel) and the OI line emission (bottom panel). The 
  dust-gas collisions are represented by the dashed curves in both panels. 
  The dotted curves show the contribution of the water molecules and the 
  dotted-dashed curves that of the CO molecules, in both panels.
  \label{heat-cool}}
\end{figure*}
Figure \ref{Tgas-profile} shows the dust and gas temperature profiles
for the Shu-like density distribution. Gas and dust are thermally
coupled across the whole envelope with the largest difference ($<$
10$\%$) in the infall region, where r $< r_{inf}$. Figure
\ref{heat-cool} shows the heating and cooling terms across the
envelope. The heating is totally dominated by the compression in the
entire collapsing region, and by the collisions between the dust and
the gas in the outer envelope. Note that the H$_2$O and CO
photo-pumping plays only a minor role. The cooling is dominated by the
water line emission in the inner region (where the ice sublimate), by
the dust-gas collisions in a large intermediate region, and by the CO
line emission in the very outer envelope. These results are very
similar to those of \citet{Cec00b}. Note that low-lying water
  lines could possibly be contaminated by the outflow driven by the
  source This effect could possibly lead to an overestimation of the
  outer water abundance. Since the ISO water lines are spectrally
  unresolved,  it is not possible to address this question using the
  ISO data.

Tables \ref{predictions_PACS} and \ref{predictions_HIFI} list the
predicted fluxes of the water lines which will be observable with HSO,
for the best fit model (H$_2$O abundance in the outer
envelope equal to $2.5\times10^{-7}$) with H$_2$O abundance in the inner
envelope equal to $0.5\times10^{-7}$, but also for the case with a larger H$_2$O inner
envelope abundance ($2\times10^{-6}$).

\section{Discussion and conclusions}\label{sec:conclusions}

The new analysis of the single dish and interferometric continuum
observations of the envelope of IRAS16293 confirms that an envelope of
about 2 M$_\odot$ surrounds the proto-binary system of
IRAS16293. The envelope can be described with a Shu-like density
distribution, corresponding to the gas collapsing towards a 2 M$_\odot$
central star. The luminosity of IRAS16293 has been re-evaluated to be
22 L$_\odot$ for a distance of 120 pc.

Both the single dish and interferometric data can be reproduced by an
envelope with an inner radius between 20 and 30 AU, equivalent to
about 0.4$''$, and smaller than the radius at which the dust
temperature reaches 100 K (the ice mantle sublimation temperature,
namely 75 to 85 AU). We found that our new analysis can reproduce the
full SED, including the Spitzer MIR data, without the necessity of a
central cavity of 800 AU radius \citep{Jor05}. The difference between
our models and the previous ones (based on the \citealt{Sch02} initial
model) is the larger bolometric luminosity (kept as a free parameter
in our models) and the lower optical thickness of the envelope (
$\tau_{100\mu m}$=2, namely twice smaller than in the Sch{\"o}ier's
models). These differences contribute to make that the predicted MIR
radiation flux agrees with the observed one. Finally, the
interferometric data, being dominated by the two components of the
binary system, do not provide significant constraints on the envelope
structure, with one exception. They exclude the case of an envelope
with a single power-law density profile centered on the mid-way point between the two
sources. Based on this assumption, \citet{Sch04} suggested the
presence of a cavity 800 AU in diameter. Since no data constrain where
the center of the envelope is located, we favor the solution with the
envelope centered on one of the two sources. In addition, the Shu-like
model fits slightly better the PDBI data and is thus our preferred
solution.

As already noted by \citet{Cec00b}, the ISO data do not allow to
constrain the inner H$_2$O abundance because the detected lines are
optically thick and cover a relatively low range of upper-level energies
($\leq 300$ cm$^{-1}$). Also the outer-envelope water abundance is relatively
poorly constrained. In addition, the relatively low spectral resolution of ISO does not allow to determine whether some lines are contaminated by the emission  from the outflow. The future observations with the HIFI spectrometer
aboard the Herschel Space Observatory, launched in May 2009, will
certainly constrain better the water abundance profile across the
IRAS16293 envelope, helping to understand the distribution of water in
protostars similar to the Sun progenitor. On the one hand, the water
abundance in the outer envelope (0.7--20$\times$10$^{-7}$) derived from
the ISO observations here is consistent with some previous estimates
of water abundance in cold gas \citep[e.g.][]{Cer97} but only
marginally consistent with other low estimates \citep[e.g.][]{Sne00},
so the new Herschel/HIFI observations, with their largely improved
spatial and spectral resolution, will be crucial in settling the
question.  On the other hand, the inner envelope abundance
($<$2$\times$10$^{-6}$) is lower than expected if all ice in the
mantles sublimates \citep[see][]{Cec00b}. Also in this case, the
Herschel/HIFI observations will help to understand this point.

Finally, as stated in the Introduction, the major scope of the present
work is to provide as accurate as possible estimates of the dust and
gas temperature profiles of the cold envelope of IRAS16293 and its
warm inner component, also known as the hot corino, to interpret the
data observed in two large projects, TIMASSS and CHESS (see
Introduction).  We are aware that the proposed description has the
intrinsic and clear limit of not taking into account the multiple
nature of the IRAS16293 system. But it has the merit in allowing the
interpretation of the single-dish observations of the upcoming
projects, within this limitation. A more detailed analysis will only be
possible once the relevant molecular emission is observed with
interferometers, resolving the two components of the system.
In absence of that, the analysis based on
a single warm component and cold envelope is the only vialable and
allows a first understanding of the chemical composition of a system
which eventually will form a star and planetary system like our own.

\begin{acknowledgements}
  We warmly thank Laurent Loinard for the very useful discussions on
  the interferometric data of IRAS16293. We acknowledge the financial
  support by PPF and the Agence Nationale pour la Recherche (ANR),
  France (contract ANR-08-BLAN-0225). The CSO is supported by the
  National Science Foundation, award AST-0540882.
\end{acknowledgements}

\bibliographystyle{aa}
\bibliography{bib}
%

\Online
\setcounter{table}{5}
\begin{table*}[h]
  \centering
  \begin{tabular}{|cccc|}
    \hline
HIFI range     &   & X$_{in}$(H$_2$O)=$2\times10^{-6}$ & X$_{in}$(H$_2$O)=$5\times10^{-8}$    \\ \hline
   Transition  & Frequency  & Flux                    & Flux     \\
               & (GHz)      & (K Km s$^{-1}$)         & (K Km s$^{-1}$) \\ \hline
  1$_{ 1 0} \rightarrow $ 1$_{ 0 1}$  &       557.0 &   32 &   32 \\
  5$_{ 3 2} \rightarrow $ 4$_{ 4 1}$  &       620.7 &  0.31 &  0 \\
  2$_{ 1 1} \rightarrow $ 2$_{ 0 2}$  &       752.0 &   23 &   23 \\
  4$_{ 2 2} \rightarrow $ 3$_{ 3 1}$  &       916.2 &  0.04 &  0.04 \\
  2$_{ 0 2} \rightarrow $ 1$_{ 1 1}$  &       987.9 &   37 &   37 \\
  3$_{ 1 2} \rightarrow $ 3$_{ 0 3}$  &      1097.3 &   26 &   26 \\
  1$_{ 1 1} \rightarrow $ 0$_{ 0 0}$  &      1113.4 &   46 &   46 \\
  7$_{ 2 5} \rightarrow $ 8$_{ 1 8}$  &      1146.6 &  0.04 &  0 \\
  3$_{ 1 2} \rightarrow $ 2$_{ 2 1}$  &      1153.1 &   34 &   34 \\
  6$_{ 3 4} \rightarrow $ 5$_{ 4 1}$  &      1158.3 &  1.0 &  0 \\
  3$_{ 2 1} \rightarrow $ 3$_{ 1 2}$  &      1162.9 &   13 &   11 \\
  4$_{ 2 2} \rightarrow $ 4$_{ 1 3}$  &      1207.6 &  1.1 &  1.1 \\
  2$_{ 2 0} \rightarrow $ 2$_{ 1 1}$  &      1228.8 &  7.6 &  7.6 \\
  5$_{ 2 3} \rightarrow $ 5$_{ 1 4}$  &      1410.7 &  5.7 &  2.1 \\
  6$_{ 4 3} \rightarrow $ 7$_{ 1 6}$  &      1574.2 &  0.10 &  0 \\
  4$_{ 1 3} \rightarrow $ 4$_{ 0 4}$  &      1602.2 &  2.2 &  2.2 \\
  2$_{ 2 1} \rightarrow $ 2$_{ 1 2}$  &      1661.0 &   40 &   37 \\
  2$_{ 1 2} \rightarrow $ 1$_{ 0 1}$  &      1669.9 &   66 &   67 \\
  4$_{ 3 2} \rightarrow $ 5$_{ 0 5}$  &      1713.9 &  2.1 &  0.1 \\
  3$_{ 0 3} \rightarrow $ 2$_{ 1 2}$  &      1716.8 &   56 &   56 \\
  7$_{ 3 4} \rightarrow $ 7$_{ 2 5}$  &      1797.2 &  2.2 &  0.1 \\
  5$_{ 3 2} \rightarrow $ 5$_{ 2 3}$  &      1867.7 &  4.6 &  0.9 \\
  6$_{ 3 4} \rightarrow $ 7$_{ 0 7}$  &      1880.8 &  0.41 &  0 \\
  8$_{ 4 5} \rightarrow $ 7$_{ 5 2}$  &      1884.9 &  0.09 &  0 \\
  3$_{ 3 1} \rightarrow $ 4$_{ 0 4}$  &      1893.7 &  0.02 &  0.02 \\
 &  &  &  \\ \hline
  \end{tabular}
  \caption{Predictions of the line fluxes (after subtraction of the continuum) 
    of the water lines observable with the Herschel spectrometer HIFI.
    The third (fourth) column reports the predictions computed assuming that the inner H$_2$O abundance 
    is equal to $2\times10^{-6}$ ($5\times10^{-8}$) and the outer H$_2$O
    abundance is $2.5\times10^{-7}$, for the Shu-like model best fit.
    \label{predictions_HIFI}}
\end{table*}

\onllongtab{7}
{\begin{longtable}{|ccccc|}
    \hline
    PACS range &    & X$_{in}$(H$_2$O)=$2\times10^{-6}$ & X$_{in}$(H$_2$O)=$5\times10^{-8}$   &    \\ \hline
   Transition  & Wavelength & Flux                    &  Flux             &  Flux      \\
               & ($\mu$m)   & (10$^{-12}$erg s$^{-1}$ cm$^{-2}$) &  (10$^{-12}$erg s$^{-1}$ cm$^{-2}$)      & (10$^{-12}$erg s$^{-1}$ cm$^{-2}$)  \\ \hline
  9$_{ 3 6} \rightarrow $ 8$_{ 4 5}$  &       62.42 &  0.05 &  0 & \\
  9$_{ 1 8} \rightarrow $ 9$_{ 0 9}$  &       62.93 &  0.4 &  0 & \\
  8$_{ 1 8} \rightarrow $ 7$_{ 0 7}$  &       63.32 &  1.0 &  0.1 & \\
  6$_{ 6 1} \rightarrow $ 6$_{ 5 2}$  &       63.91 &  0.2 &  0 & \\
  7$_{ 6 1} \rightarrow $ 7$_{ 5 2}$  &       63.96 &  0.05 &  0 & \\
  6$_{ 2 5} \rightarrow $ 5$_{ 1 4}$  &       65.17 &  1.7 &  0.5 & \\
  7$_{ 1 6} \rightarrow $ 6$_{ 2 5}$  &       66.09 &  1.2 &  0.2 & \\
  3$_{ 3 0} \rightarrow $ 2$_{ 2 1}$  &       66.44 &  2.5 &  1.6 & \\
  3$_{ 3 1} \rightarrow $ 2$_{ 2 0}$  &       67.09 &  0.3 &  0.3 & \\
  3$_{ 3 0} \rightarrow $ 3$_{ 0 3}$  &       67.27 &  2.0 &  0.5 & \\
  8$_{ 2 7} \rightarrow $ 8$_{ 1 8}$  &       70.70 &  0.6 &  0 & \\
  5$_{ 2 4} \rightarrow $ 4$_{ 1 3}$  &       71.07 &  0.03 &  0.03 & \\
  7$_{ 0 7} \rightarrow $ 6$_{ 1 6}$  &       71.95 &  1.1 &  0.2 & \\
  7$_{ 2 5} \rightarrow $ 6$_{ 3 4}$  &       74.95 &  0.6 &  0 & \\
  3$_{ 2 1} \rightarrow $ 2$_{ 1 2}$  &       75.38 &  2.9 &  2.4 & 2.7$\pm$1.0 \\
  6$_{ 5 2} \rightarrow $ 6$_{ 4 3}$  &       75.83 &  0.3 &  0 & \\
  5$_{ 5 0} \rightarrow $ 5$_{ 4 1}$  &       75.91 &  0.5 &  0 & \\
  7$_{ 5 2} \rightarrow $ 7$_{ 4 3}$  &       77.76 &  0.2 &  0 & \\
  4$_{ 2 3} \rightarrow $ 3$_{ 1 2}$  &       78.74 &  2.0 &  1.5 & \\
  9$_{ 2 7} \rightarrow $ 9$_{ 1 8}$  &       81.41 &  0.10 &  0 & \\
  6$_{ 1 6} \rightarrow $ 5$_{ 0 5}$  &       82.03 &  1.1 &  0.5 & \\
  8$_{ 3 6} \rightarrow $ 8$_{ 2 7}$  &       82.98 &  0.2 &  0 & \\
  6$_{ 0 6} \rightarrow $ 5$_{ 1 5}$  &       83.28 &  0.04 &  0.04 & \\
  7$_{ 1 6} \rightarrow $ 7$_{ 0 7}$  &       84.77 &  0.8 &  0.1 & \\
  8$_{ 4 5} \rightarrow $ 8$_{ 3 6}$  &       85.77 &  0.06 &  0 & \\
  3$_{ 2 2} \rightarrow $ 2$_{ 1 1}$  &       89.99 &  0.5 &  0.5 &  0.5$\pm$0.5 \\
  6$_{ 4 3} \rightarrow $ 6$_{ 3 4}$  &       92.81 &  0.3 &  0 & \\
  6$_{ 2 5} \rightarrow $ 6$_{ 1 6}$  &       94.64 &  0.7 &  0.2 & \\
  4$_{ 4 1} \rightarrow $ 4$_{ 3 2}$  &       94.71 &  0.6 &  0.1 & \\
  5$_{ 1 5} \rightarrow $ 4$_{ 0 4}$  &       95.63 &  0.3 &  0.3 & \\
  5$_{ 4 1} \rightarrow $ 5$_{ 3 2}$  &       98.49 &  0.5 &  0 & \\
  5$_{ 0 5} \rightarrow $ 4$_{ 1 4}$  &       99.49 &  1.3 &  1.1 & \\
  5$_{ 1 4} \rightarrow $ 4$_{ 2 3}$  &      100.91 &  1.1 &  0.8 &    1.3$\pm$0.6 \\
  2$_{ 2 0} \rightarrow $ 1$_{ 1 1}$  &      100.98 &  0.8 &  0.8 & \\
  6$_{ 3 4} \rightarrow $ 6$_{ 2 5}$  &      104.09 &  0.3 &  0 & \\
  2$_{ 2 1} \rightarrow $ 1$_{ 1 0}$  &      108.07 &  2.2 &  2.1 &   1.7$\pm$0.6 \\
  7$_{ 4 3} \rightarrow $ 7$_{ 3 4}$  &      112.51 &  0.07 &  0 & \\
  4$_{ 1 4} \rightarrow $ 3$_{ 0 3}$  &      113.54 &  1.5 &  1.3 & \\
  7$_{ 3 4} \rightarrow $ 6$_{ 4 3}$  &      116.78 &  0.10 &  0 & \\
  4$_{ 3 2} \rightarrow $ 4$_{ 2 3}$  &      121.72 &  0.4 &  0.2 & \\
  4$_{ 0 4} \rightarrow $ 3$_{ 1 3}$  &      125.36 &  0.6 &  0.6 & \\
  3$_{ 3 1} \rightarrow $ 3$_{ 2 2}$  &      126.71 &  0.1 &  0.1 & \\
  7$_{ 2 5} \rightarrow $ 7$_{ 1 6}$  &      127.88 &  0.10 &  0 & \\
  4$_{ 2 3} \rightarrow $ 4$_{ 1 4}$  &      132.41 &  0.7 &  0.6 & 0.7$\pm$0.7 \\
  8$_{ 3 6} \rightarrow $ 7$_{ 4 3}$  &      133.55 &  0.03 &  0 & \\
  5$_{ 1 4} \rightarrow $ 5$_{ 0 5}$  &      134.94 &  0.5 &  0.3 & \\
  3$_{ 3 0} \rightarrow $ 3$_{ 2 1}$  &      136.49 &  0.6 &  0.4 & \\
  3$_{ 1 3} \rightarrow $ 2$_{ 0 2}$  &      138.53 &  0.9 &  0.9 & 0.9$\pm$0.7 \\
  4$_{ 1 3} \rightarrow $ 3$_{ 2 2}$  &      144.52 &  0.1 &  0.1 & \\
  3$_{ 2 2} \rightarrow $ 3$_{ 1 3}$  &      156.20 &  0.2 &  0.2 & \\
  5$_{ 2 3} \rightarrow $ 4$_{ 3 2}$  &      156.26 &  0.3 &  0.04 & \\
  5$_{ 3 2} \rightarrow $ 5$_{ 2 3}$  &      160.51 &  0.1 &  0 & \\
  7$_{ 3 4} \rightarrow $ 7$_{ 2 5}$  &      166.81 &  0.05 &  0 & \\
  3$_{ 0 3} \rightarrow $ 2$_{ 1 2}$  &      174.62 &  1.2 &  1.2 &  2.5$\pm$0.6 \\
  4$_{ 3 2} \rightarrow $ 5$_{ 0 5}$  &      174.92 &  0.05 &  0 & \\
  2$_{ 1 2} \rightarrow $ 1$_{ 0 1}$  &      179.53 &  1.4 &  1.4 &  2.9$\pm$0.6\\
  2$_{ 2 1} \rightarrow $ 2$_{ 1 2}$  &      180.49 &  0.9 &  0.8 & 0.9$\pm$0.4 \\
  4$_{ 1 3} \rightarrow $ 4$_{ 0 4}$  &      187.11 &  0.05 &  0.05 & \\
\\ \hline
  \caption{Predictions of the line fluxes (after subtraction of the continuum) 
    of the water lines observable with the Herschel spectrometer PACS.
    The third (fourth) column reports the predictions computed assuming that the inner H$_2$O abundance 
    is equal to $2\times10^{-6}$ ($5\times10^{-8}$) and the outer H$_2$O
    abundance is $2.5\times10^{-7}$, for the Shu-like model best fit. The fifth column reports the observed ISO fluxes
    \label{predictions_PACS}}

\end{longtable}
}


\end{document}